\def\rfr#1{eq. (\ref{#1})}

\def\dert#1#2{\frac{{{d}}{#1}}{{{d}}{#2}}}              

\def\virg#1{``#1''}

\def\eqi{\begin{equation}}
\def\eqf{\end{equation}}
\def\eqia{\begin{eqnarray}}
\def\eqfa{\end{eqnarray}}
\def\rp#1#2{{#1\over#2}} \def\lb#1{\label{#1}}

\documentclass[11pt]{article}
\usepackage{amsmath,amsthm,amscd,amssymb,wasysym}
\usepackage{latexsym}
\usepackage{graphicx,epsfig}
\usepackage{abstract}
\RequirePackage{color}

\begin{document}

	
\noindent{\bf \LARGE{Solar system constraints on a Rindler-type extra-acceleration from modified gravity at large distances}}
\\
\\
\\
{L. Iorio$^{\ast}$}\\
{\it $^{\ast}$Ministero dell'Istruzione, dell'Universit\`{a} e della Ricerca (M.I.U.R.)\\ Fellow of the Royal Astronomical Society (F.R.A.S.)\\
 Permanent address: Viale Unit$\grave{a}$ di Italia 68
70125 Bari (BA), Italy.  \\ e-mail: lorenzo.iorio@libero.it}

\begin{onecolabstract}
We analytically work out the orbital effects caused by a Rindler-type extra-acceleration $A_{\rm Rin}$ which naturally arises in some recent models of modified gravity at large distances. In particular, we focus on the perturbations induced by it on the two-body range $\rho$ and range-rate $\dot\rho$ which are commonly used in satellite and planetary investigations as primary observable quantities.
The constraints obtained for $A_{\rm Rin}$ by comparing our calculations with the currently available range and range-rate residuals for some of the major bodies of the solar system, obtained without explicitly modeling $A_{\rm Rin}$, are  $1-2\times 10^{-13}$ m s$^{-2}$ (Mercury and Venus), $1\times 10^{-14}$ m s$^{-2}$ (Saturn), $1\times 10^{-15}$ m s$^{-2}$ (Mars), while for a terrestrial Rindler acceleration we have an upper bound of $5\times 10^{-16}$ m s$^{-2}$ (Moon). \textcolor{black}{The  constraints inferred from the planets' range and range-rate residuals are confirmed also by the latest empirical determinations of the corrections $\Delta\dot\varpi$ to the usual Newtonian/Einsteinian secular precessions of the planetary longitudes of perihelia $\varpi$: moreover, the Earth yields $A_{\rm Rin}\leq 7\times 10^{-16}$ m s$^{-2}$.}
Another approach which could be followed consists of taking into account $A_{\rm Rin}$ in re-processing all the available data sets with accordingly modified dynamical models, and estimating a dedicated solve-for parameter explicitly accounting for it. Anyway, such a method is time-consuming. A preliminary analysis likely performed in such a way by a different author yields $A\leq 8\times 10^{-14}$ m s$^{-2}$ at  Mars' distance and $A\leq 1\times 10^{-14}$ m s$^{-2}$ at Saturn's distance.
The method adopted here can be easily  and straightforwardly extended to other long-range modified models of gravity as well.
\end{onecolabstract}
Keywords: Experimental tests of gravitational theories, Modified theories of gravity, Celestial mechanics, Ephemerides, almanacs, and calendars, Orbit determination and improvement   \\ \\
PACS numbers: 04.80.Cc, 04.50.Kd,  95.10.Ce, 95.10.Km, 95.10.Eg\\


\section{Introduction}
Recently, Grumiller \cite{gru} has constructed an effective model for gravity of a central object of mass $M$ at large scales. Starting from
the most general space-time metric endowed with spherical symmetry in four dimensions \cite{wald}
\eqi (ds)^2 = g_{AB}(x^{\mu})dx^A dx^B +\Phi^2(x^{\mu})\left[\left(d\theta\right)^2 +\sin^2\theta \left(d\phi\right)^2\right], A,B=0,1,\eqf where $\Phi(x^{\mu})$ is a 2-dimensional dilaton field, it is possible to reduce the 4-dimensional Einstein-Hilbert action to a 2-dimensional dilaton\footnote{Dilaton gravity models in two dimensions have recently attracted much attention because of their implications for a list of well-known problems in quantum gravity \cite{gru2}.} one \cite{gru2}.
By writing down the most general 2-dimensional action for $g_{AB}$ and $\Phi$ consistent with spherical symmetry and with additional assumptions like power-counting renormalizability, analyticity, etc., and by considering the long-distance limit, an additional term with respect to the usual ones of the General Theory of Relativity (GTR) appears. It is \cite{gru}
\eqi g_{AB}(x^{\mu})dx^A dx^B = K^2 (dx^0)^2 -\rp{(dr)^2}{K^2},\ \Phi =r,\eqf
with
\eqi K^2 = 1-\rp{2GM}{r}-\Lambda r^2 + 2 \mathcal{A}r. \eqf
While $M$ and $\Lambda$ are parameters retaining their usual GTR meanings,  $\mathcal{A}$, according to its geometrical interpretation \cite{gru2}, yields a radially directed Rindler extra-acceleration $A_{\rm Rin}\doteq \mathcal{A}c^2$ (RIN in the following) \cite{wald}.

Since in Ref.~\cite{gru} it has been envisaged the possibility that RIN may explain the anomalous behavior exhibited by both the Pioneer spacecrafts after they passed the threshold of about 20 astronomical unit (au) \cite{pio1,pio2}, traditionally expressed in terms of an unexplained radial acceleration \eqi A_{\rm Pio}=(8.74\pm 1.33)\times 10^{-10}\ {\rm m\ s}^{-2},\lb{apio}\eqf it is worthwhile to study in more details the phenomenology of RIN in the solar system. Incidentally, let us recall that recent works \cite{piong1,piong2} point towards an explanation of the Pioneer anomaly in terms of mundane, non-gravitational effects peculiar to the probes. On the other hand, if the Pioneer anomaly  was really a genuine dynamical effect of gravitational origin, it should
also affect the orbital motions of the solar system's bodies moving in the space regions in
which it manifested itself in its presently known form. For such an issue see, e.g., Ref.~\cite{iorio} and references therein.

In Section \ref{anale} we analytically work out some effects caused by RIN  on test particles' orbital motion around a central body. In particular, in Section \ref{Rrange} we deal with the two-body range, while in Section \ref{Rrange_rate} we treat its time derivative, i.e. the two-body range-rate. In Section \ref{residui} we put constraints on RIN by comparing our results with the most recently produced range and range-rate residuals for several solar system bodies. Section \ref{conclusioni} is summarizes our findings.

Finally, let us note that the strategy presented here can, in principle,  be straightforwardly extended to other exotic effects predicted by different long-range modified models of gravity.
\section{Analytical calculation}\lb{anale}
In principle, simple back-to-the-envelope computations may be performed by noting that, given a constant and uniform extra-acceleration $A_{\rm Rin} $ acting on a test particle in orbital motion around a central body, its position and velocity shifts caused by $A_{\rm Rin} $ are roughly given by the product of $A_{\rm Rin} $ by the second and first powers, respectively, of a characteristic time $T$ of the system considered which, in the present case, is the particle's orbital period $P_{\rm b}$. Anyway, such a naive approach is not able to tell us if such a kind of perturbation does actually affect the orbital motions with non-vanishing, long-term effects. Moreover, also by a-priori supposing that it is just the case, the correct order of magnitude of them may not be correctly inferred because of the true details of the orbit like, e.g., its eccentricity $e$. Indeed, since it is usually very small for typical solar system bodies, its presence or its absence in the expression for a certain orbital effect may substantially alter its  size. Thus, it is mandatory to explicitly work out in full details the perturbations induced by $A_{\rm Rin} $ on some features of test particles' orbits: we will choose the\footnote{We refer to two test particles A and B orbiting the same central body and reciprocally interconnected by means of some artificial, man-made  electromagnetic links.} two-body range $\rho$ and  range-rate $\dot\rho$ because they are direct, unambiguous and accurate observables very common in Earth-satellite and Sun-planets investigations. In Section \ref{Rrange} the perturbation  $\Delta\rho$ is computed, while Section \ref{Rrange_rate} is devoted to $\Delta\dot\rho$.
\subsection{The two-body range}\lb{Rrange}
The orbit of a test particle around a central mass in presence of a dynamical perturbation $\vec{A}$ of the usual two-body, pointlike Newtonian monopole can be written as
\eqi\vec{r}_{\rm P}=\vec{r}_{\rm U}+\Delta\vec{r}.\eqf
The subscript\footnote{In the following we will neglect it to make the notation less cumbersome.} P denotes the perturbed orbit, while U labels the unperturbed Keplerian ellipse. \textcolor{black}{To avoid  possible misunderstandings, we remark that what will, actually, be compared to the data in Section \ref{residui} is only obtained from the perturbation $\Delta \vec{r}$.} \textcolor{black}{As a consequence, standard perturbative approaches like those relying upon the Gauss or the Lagrange equations  for the variation of the Keplerian orbital elements \cite{Roy} can be adopted to work out the orbital effects of $\vec{A}$. They make use of the Newtonian trajectory as unperturbed, reference path.
In principle, it is possible to assume as reference orbit a fully post-Newtonian one \cite{Calura1,Calura2}, and work out the effects of a given  small extra-acceleration $\vec{A}$ with respect to it according to the perturbative scheme set up by the authors of Refs. \cite{Calura1,Calura2}. It is a general relativistic generalization of the standard perturbative approach based on the planetary Lagrange equations \cite{Roy}. In the present case, given the extremely tight bounds on $A_{\rm Rin}$ which will be inferred in Section \ref{residui}, we may, in principle, safely apply such relativistic perturbative scheme since $A_{\rm Rin}$ will turn out to be not only much smaller than the Newtonian monopole term, but also of the Schwarzschild-like planetary accelerations which are of the order of
\eqi A_{\rm GTR}\approx \rp{(GM)^2}{c^2 r^3}= 1\times 10^{-9}-6\times 10^{-11}\ {\rm m\ s^{-2}}.\eqf
Anyway, it would be, in practice, useless since the only addition with respect to the orbital effects resulting from the standard perturbative scenario would consist of further, small cross GTR-RIN orbital effects which, in the present case, would be  of the order of $A_{\rm Rin}(v/c)^2$. Since typical planetary speeds are of the order of $v\approx\sqrt{GM/r}\approx 10^4$ m s$^{-1}$, such mixed terms would be completely negligible.
} {Let us remark that putting a bound smaller than the GTR Schwarzschild-type term is, actually, consistent. Indeed,  GTR is fully modelled in the softwares routinely used to process astronomical data: thus, what actually is relevant is the uncertainty in the GTR terms, which is, at present, at a $0.01$ percent level or less (see Section \ref{rangi} below). Moreover, as it will be shown in Section \ref{rangi}, the RIN and GTR effects have also different time signatures, thus allowing for  tighter constraining of RIN.  }

For analytical purposes, we will work in the $R-T-N$ formalism by expressing the position perturbation $\Delta\vec{r}$  in terms of its radial, transverse and out-of-plane (or normal) projections $\Delta R,\Delta T,\Delta N$ onto the three orthogonal directions of the co-moving frame with unit vectors $\hat{r},\hat{\tau},\hat{\nu}$, so that
\eqi\Delta \vec{r}=\Delta R\ \hat{r}+\Delta T\ \hat{\tau}+\Delta N\ \hat{\nu}.\lb{RTN}\eqf

In order to analytically work out the range and range-rate perturbations due to some dynamical effects, let us review in some details some key-features of the $R-T-N$ formalism.
In terms of the standard Keplerian orbital elements and of the usual unit vectors $\hat{\imath},\hat{\jmath},\hat{k}$ of an inertial frame with Cartesian rectangular  coordinates having its origin in the central body, the $R-T-N$ versors, evaluated onto the unperturbed orbit, are \cite{Mont}
\eqi \hat{{r}} =\left(
       \begin{array}{c}
          \cos\Omega\cos u\ -\cos I\sin\Omega\sin u\\
          \sin\Omega\cos u + \cos I\cos\Omega\sin u\\
         \sin I\sin u \\
       \end{array}
     \right)\lb{krizza}
\eqf
 \eqi \hat{{\tau}} =\left(
       \begin{array}{c}
         -\sin u\cos\Omega-\cos I\sin\Omega\cos u \\
         -\sin\Omega\sin u+\cos I\cos\Omega\cos u \\
         \sin I\cos u \\
       \end{array}
     \right)
\eqf
 \eqi \hat{{\nu}} =\left(
       \begin{array}{c}
          \sin I\sin\Omega \\
         -\sin I\cos\Omega \\
         \cos I,\\
       \end{array}
     \right)
\eqf
where $I,\Omega, \omega$ are the inclination of the orbit to the reference $\{x,y\}$ plane adopted, the longitude of the ascending node and the argument of the pericentre, respectively; $u\doteq \omega+f$ is the argument of latitude, in which $f$ is the true anomaly reckoning the instantaneous position of the test particle along its Keplerian ellipse.
Thus, after having analytically  worked out the perturbations of the Keplerian orbital elements, it will be possible to calculate $\Delta R,\Delta T, \Delta N$ according to Ref.~\cite{Caso}
\begin{equation}
\left\{\begin{array}{lll}
\Delta R &=&\left(\rp{r}{a}\right)\Delta a - a \cos f\Delta e +\rp{ae\sin f}{\sqrt{1-e^2}}\Delta{\mathcal{M}}, \\ \\
\Delta T &=& a\sin f\left[1+\rp{r}{a(1-e^2)}\right]\Delta e + r(\cos I\Delta\Omega + \Delta\omega)+\left(\rp{a^2}{r}\right)\sqrt{1-e^2}\Delta{\mathcal{M}},\\ \\
\Delta N &=& r(\sin u\Delta I-\cos u\sin I\Delta\Omega),
\end{array}\lb{dnorm}
\right.
\end{equation}
in which $a,e,{\mathcal{M}}$ are the semimajor axis, the eccentricity and the mean anomaly, respectively.

The Gauss equations for the variation of the Keplerian orbital elements are \cite{Roy}
\begin{equation}
\left\{
\begin{array}{lll}
\dert a t & = & \rp{2}{n\sqrt{1-e^2}} \left[e A_R\sin f +A_{T}\left(\rp{p}{r}\right)\right],\\   \\
\dert e t  & = & \rp{\sqrt{1-e^2}}{na}\left\{A_R\sin f + A_{T}\left[\cos f + \rp{1}{e}\left(1 - \rp{r}{a}\right)\right]\right\},\\  \\
\dert I t & = & \rp{1}{na\sqrt{1-e^2}}A_N\left(\rp{r}{a}\right)\cos u,\\   \\
\dert\Omega t & = & \rp{1}{na\sin I\sqrt{1-e^2}}A_N\left(\rp{r}{a}\right)\sin u,\\    \\
\dert\omega t & = &\rp{\sqrt{1-e^2}}{nae}\left[-A_R\cos f + A_{T}\left(1+\rp{r}{p}\right)\sin f\right]-\cos I\dert\Omega t,\\   \\
\dert {\mathcal{M}} t & = & n - \rp{2}{na} A_R\left(\rp{r}{a}\right) -\sqrt{1-e^2}\left(\dert\omega t + \cos I \dert\Omega t\right).
\end{array}\lb{Gauss}
\right.
\end{equation}
 In \rfr{Gauss} $p\doteq a(1-e^2)$ is the semi-latus rectum, $n\doteq\sqrt{GM/a^3}$ is the unperturbed Keplerian mean motion related to the Keplerian orbital period by $n=2\pi/P_{\rm b}$, and $A_R,A_T,A_N$ are the radial, transverse and out-of-plane components of the disturbing acceleration $\vec{A}$ which have to be computed onto the unperturbed Keplerian ellipse.
 It turns out that, in order to make the calculations easier, it is more convenient to use the eccentric anomaly $E$ instead of the true anomaly $f$; basically, $E$ can be regarded as
a parametrization of the polar angle in the orbital plane. To this aim, useful conversion relations are \cite{Roy}
 \eqi
 \left\{
 \begin{array}{lll}
 \cos f &=& \rp{\cos E-e}{1-e\cos E}, \\ \\
 \sin f &=& \rp{\sqrt{1-e^2}\sin E}{1-e\cos E},\\ \\
 r &=& a(1-e\cos E), \\ \\
 dt &=& \left(\rp{1-e\cos E}{n}\right)d E.
 \end{array}\lb{conz}
 \right.
 \eqf
%
%

In the specific case of RIN, the $R-T-N$ components of its acceleration are simply
\begin{equation}
\left\{\begin{array}{lll}
A_R &=& A_{\rm Rin} ,\\  \\
A_T &=& 0,\\  \\
A_N &=& 0,
\end{array}\lb{arelt}
\right.
\end{equation}
in which $A_{\rm Rin} $ can be either positive or negative.
 Inserting \rfr{arelt} into \rfr{Gauss} and integrating it by means of \rfr{conz} from  the initial value of the eccentric anomaly $E_0$ to a subsequent, generic value $E$ yield
\begin{equation}
\left\{
\begin{array}{lll}
\Delta a & = & -\rp{2eA_{\rm Rin} \left(\cos E-\cos E_0\right)}{n^2}, \\ \\
\Delta e & = & -\rp{A_{\rm Rin} \left(1-e^2\right)\left(\cos E-\cos E_0\right)}{n^2},\\  \\
\Delta I & = & 0, \\  \\
\Delta \Omega & = &  0,\\ \\
\cos I\Delta \Omega +\Delta\omega & = & \rp{A_{\rm Rin} \sqrt{1-e^2}}{an^2} \left[\left(E-E_0\right)-\rp{\left(\sin E-\sin E_0\right)}{e}\right], \\  \\
\Delta {\mathcal{M}}& = & \rp{A_{\rm Rin} }{an^2}\left[-3\left(E-E_0\right) +\rp{\left(\sin E-\sin E_0\right)}{e}+\right.\\ \\
&+&\left. 3e\left(\sin E-\sin E_0\right)-\rp{e^2\left(\sin 2 E-\sin 2 E_0 \right)}{2}\right].
\end{array}\lb{equaz}
\right.
\end{equation}

Thus, \rfr{equaz}  inserted into \rfr{dnorm} yield the $R-T-N$ position perturbations due to  RIN
\eqi
\left\{
\begin{array}{lll}
  \Delta R &=& -\rp{A_{\rm Rin} }{n^2\left(1-e\cos E\right)}\left\{\right.\\ \\
  & &\left.\left[\cos \left(E-E_0\right)-1\right] +\right. \\ \\
    & + &\left. 3e\left[(E-E_0)\sin E +\cos E-\cos E_0\right]+\right.\\ \\
    &+&\left. 3e^2\left[\cos(E-E_0)-1\right]+\right. \\ \\
    &+&\left.\rp{e^3}{4}\left[3\cos E+\cos 3 E-4\cos E_0\left(\cos 2 E+\sin E\sin E_0\right)\right]\right.\\ \\
    & &\left.  \right\},\\  \\
  \Delta T &=&  -\rp{A_{\rm Rin} }{n^2\left(1-e\cos E\right)}\left\{\right.\\  \\
  & & \left. 2\left[\sin\left(E-E_0\right) -\left(E-E_0\right)\right] +\right. \\ \\
  &+&\left. \rp{e}{2}\left[4\left(E-E_0\right)\cos E -6\sin E +\sin\left(2E-E_0\right) +5\sin E_0 \right]-\right. \\ \\
  &-& \left. e^2\left[(E-E_0)\cos^2 E  -\left(\sin E-\sin E_0\right)\cos E_0\right]\right.\\ \\
  & &\left. \right\}, \\ \\
  \Delta N &=& 0.
\end{array}\lb{inculescion}
\right.
\eqf
Note that the results of \rfr{equaz} and of \rfr{inculescion} are exact; no approximations in $e$ have been used. Moreover, \rfr{inculescion} does not present any singularities for particular values of  $e$.
%
%
%

In order to conveniently plot \rfr{inculescion} as a function of time we will use suitable  partial sums of the series\footnote{It converges for all $e<1$ like a geometric series with ratio ${\rm r}=\left[e/\left(1+\sqrt{1-e^2}\right)\right]\exp\left(\sqrt{1-e^2}\right)$ \cite{series}. See also  http://mathworld.wolfram.com/KeplersEquation.html.}
\eqi
E=\mathcal{M}+\sum_{q=1}^{\rm \infty}\rp{2}{q}J_q(qe)\sin\left( q\mathcal{M}\right),\lb{anom}
\eqf
where $J_q(qe)$ is the Bessel function of the first kind \cite{bessel}.
 Indeed, the mean anomaly is a parametrization of time according to  \eqi{\mathcal{M}}\doteq n(t-t_p),\eqf where $t_p$ is the time of the passage at pericenter.
It turns out that for small eccentricities just a few terms in \rfr{anom} have to be retained.
Note that \rfr{inculescion} tells us that, for a given value of $A_{\rm Rin} $, the largest effects occur for those orbiters having the largest orbital periods; in particular, for a given central body of mass $M$, the most distant test particles orbiting it experience the largest perturbations. Indeed, the shifts of \rfr{inculescion} are proportional to $A_{\rm Rin}P^2_{\rm b}$, as expected. This fact is important because, for a given level of accuracy in determining the orbits of the probes used, the tightest constraints on $A_{\rm Rin} $ come just from the most distant ones with respect to $M$.

The perturbations of \rfr{inculescion} can  fruitfully be used to analytically work out the two-body range and range-rate perturbations between two test particles A and B orbiting the same central body of mass $M$.
Indeed, concerning the range, from
\eqi
\left\{
\begin{array}{lll}
\rho^2 &=&\left(\vec r_{\rm A}-\vec r_{\rm B}\right)\cdot\left(\vec r_{\rm A}-\vec r_{\rm B}\right),\\ \\
\hat{\rho} &\doteq & \rp{\left(\vec r_{\rm A}-\vec r_{\rm B}\right)}{\rho},
\end{array}
\right.
\eqf
to be evaluated onto the unperturbed Keplerian ellipses
of the two test particles A and B,
it follows that, for a generic perturbation, the range shift $\Delta \rho$ is \cite{cheng}
\eqi\Delta\rho=\left(\Delta\vec r_{\rm A}-\Delta\vec r_{\rm B}\right)\cdot\hat{\rho}.\lb{range}\eqf
In the case of the unperturbed Keplerian ellipse it is
\eqi \vec r= r\ \hat{r},\eqf with $r$ as in \rfr{conz}, and $\hat{r}$ given by \rfr{krizza}.
\subsection{The two-body range-rate}\lb{Rrange_rate}
It is also possible to analytically work out  the two-body range-rate perturbation $\Delta\dot\rho$ \cite{cheng} by, first, working out the $R-T-N$ shifts of the test-particle's velocity. In general, they are
%
\cite{Caso}
\eqi
\left\{
\begin{array}{lll}
\Delta v_R &=&    -\rp{n\sin f}{\sqrt{1-e^2}} \left( \rp{e\Delta a}{2} + \rp{a^2\Delta e}{r}\right)-\rp{n a^2\sqrt{1-e^2}}{r}\left(\cos I\Delta\Omega+\Delta\omega\right)-\rp{na^3}{r^2}\Delta \mathcal{M}, \\ \\
\Delta v_T &=& -\rp{na\sqrt{1-e^2}}{2r}\Delta a +\rp{an\left(e+\cos f\right)}{(1-e^2)^{3/2}}\Delta e +\rp{nae\sin f}{\sqrt{1-e^2}}\left(\cos I\Delta\Omega+\Delta\omega\right), \\ \\
\Delta v_N &=& \rp{na}{\sqrt{1-e^2}}\left[\left(\cos u+e\cos\omega\right)\Delta I +\left(\sin u +e \sin\omega\right)\sin I \Delta\Omega\right],
\end{array}\lb{velarr}
\right.
\eqf
so that
\eqi \Delta \vec{v} = \Delta v_R\ \hat{r}+\Delta v_T\ \hat{\tau}+\Delta v_N\ \hat{\nu}.\eqf
In the case of RIN, inserting \rfr{equaz} in \rfr{velarr} straightforwardly yields
\eqi
\left\{
\begin{array}{lll}
\Delta v_R & = & \rp{A_{\rm Rin} }{n(1-e\cos E)^2}\left\{    \right.\\ \\
& & \left.  \left(E-E_0\right)\left(2+e\cos E\right)-\sin \left(E-E_0\right) -\right. \\ \\
& - & \left. e\left(4 -e\cos E\right)\left(\sin E-\sin E_0\right)+ \right. \\ \\
&+& \left. \rp{e^2}{2}\left[ e\cos E_0\sin 2 E+ \left(1-e\cos E\right)\left(2E-2E_0 +\sin2 E\right) -\sin 2 E_0 \right]\right. \\ \\
& &\left.  \right\},\\ \\
\Delta v_T & = & \rp{A_{\rm Rin} \sqrt{1-e^2}}{n\left(1-e\cos E\right)}\left[\right. \\ \\
& & \left. \left(E-E_0\right)\sin E -1 +\cos(E-E_0) +\right. \\ \\
&+& \left. e\left(\cos E-\cos E_0\right)\right.\\ \\
& & \left.\right],\\ \\
\Delta v_N & = & 0.
\end{array}\lb{pluss}
\right.
\eqf
Also \rfr{pluss} are exact in $e$. Note that, for a given value of $A_{\rm Rin} $, also the shifts of \rfr{pluss} get larger for more distant, i.e. slower, orbiting test particles around a central body; indeed, as expected, they are proportional to $A_{\rm Rin}P_{\rm b}$.

In order to work out the two-body range-rate perturbation, the following unit vector, computed onto the unperturbed Keplerian ellipse, is needed \cite{cheng}
\eqi \hat{\rho}_n\doteq \rp{\left(\vec{v}_{\rm A}-\vec{v}_{\rm B}\right)-\dot\rho\ \hat{\rho}}{\rho},\lb{vel1}\eqf
where
\eqi \dot\rho = \left(\vec{v}_{\rm A}-\vec{v}_{\rm B}\right)\cdot\hat{\rho}.\lb{vel2}\eqf
In \rfr{vel1}-\rfr{vel2}
the\footnote{Here we drop the subscript U denoting the unperturbed trajectory.} Keplerian test particle's velocity must be used; it is
\eqi \vec{v}=v_R\ \hat{r}+v_T\ \hat{\tau},\eqf
with
\eqi
\left\{
\begin{array}{lll}
v_R &=& \rp{nae\sin f}{\sqrt{1-e^2}}, \\ \\
v_T &=& \rp{na\left(1+e\cos f\right)}{\sqrt{1-e^2}}.
\end{array}
\right.
\eqf
Note that, by construction, $\hat{\rho}_n$ is orthogonal to $\hat{\rho}$.
The two-body range-rate perturbation $\Delta\dot\rho$ is, thus, \cite{cheng}
\eqi \Delta\dot\rho = \left(\Delta \vec{v}_{\rm A}-\Delta \vec{v}_{\rm B}\right)\cdot \hat{\rho} + \left(\Delta \vec{r}_{\rm A}-\Delta \vec{r}_{\rm B}\right)\cdot \hat{\rho}_n.\lb{rangerate}\eqf
In the present specific case, inserting \rfr{inculescion} and \rfr{pluss} into \rfr{rangerate}, with \rfr{anom}, allows to plot the range-rate perturbation due to RIN as a function of time.
Finally, let us note that an alternative approach to compute $\Delta\dot\rho$ consists of straightforwardly  taking the derivative $\Delta\rho$ with respect to $t$ after that its time series has been generated from \rfr{range} with the aid of \rfr{anom}.
\section{Confrontation with the observations}\lb{residui}
\textcolor{black}{\subsection{The range and range-rate residuals}\lb{rangi}}
Here we plot the analytically computed time-series of the two-body range and range-rate perturbations caused by RIN  for A$=$Earth and B given by various bodies of the solar system. Then, we compare them with the existing residuals produced with the latest ephemerides in which RIN was not modeled in order to preliminarily put constraints on the magnitude of $A_{\rm Rin} $. \textcolor{black}{In order to unambiguously and correctly interpret our results, it should be noticed that all standard Newtonian and Einsteinian gravitational effects have been accurately modeled in computing the range-residuals, so that they account, in principle,  for any unmodeled dynamical effect like just RIN}.
{In particular, the modeling of GTR dynamical effects has reached a sub-percent level of accuracy. Indeed, reasoning in terms of the PPN parameters $\beta$ and $\gamma$, latest determinations of them from fitting dynamical models of different ephemerides (INPOP, EPM, DE) to large data sets covering about one century by independent teams of astronomers are accurate at $\sim 10^{-4}-10^{-5}$ level \cite{Berto,pitjeva,folkner,Fie010,PitjevaJournees010,FiengaJournees010,FolknerJournees010,Lam, Kono}.}
We are confident in our results because we independently checked our analytically-produced time series by numerically integrating the equations of motion for the pairs A-B with and without RIN; the resulting range and range-rate signatures coincide with those analytically computed.

According to Table 1 of Ref.~\cite{pitjeva}, the Mercury range residuals cover rather continuously a time span 33 yr long (1964-1997) with a Root-Mean-Square (RMS) error of 575 m; see also Figure B-2 and Figure B-3 of Ref.~\cite{DE}. Figure \ref{mercury_range} shows that the largest admissible value for $A_{\rm Rin} $ yielding an unmodelled range signal still compatible with the existing  residuals is $10^{-13}$ m s$^{-2}$; indeed, a larger value for $A_{\rm Rin} $ would induce a signature larger than the actual residuals.
\begin{figure*}[ht!]
\centering
\begin{tabular}{c}
\epsfig{file=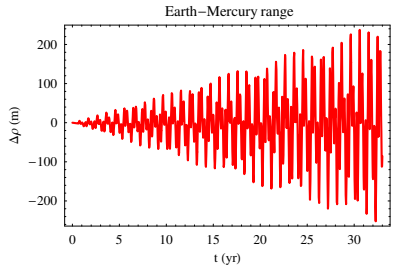,width=0.45\linewidth,clip=}
\end{tabular}
\caption{Earth-Mercury range perturbation $\Delta\rho$, in m, caused by $A_{\rm Rin} =1\times 10^{-13}$ m s$^{-2}$ over $\Delta t=33$ yr. The initial conditions have been retrieved from the NASA-JPL WEB interface HORIZONS. }\lb{mercury_range}
\end{figure*}

Table 1 of Ref.~\cite{pitjeva} tells us that the Venus range residuals span rather continuously an interval 34 yr long (1961-1995) with a RMS error of 584 m; see also Figure B-6 of Ref.~\cite{DE}. The Magellan range-rate residuals covering 2 yr (1992-1994) have a RMS error of just $0.007$ mm s$^{-1}$ \cite{pitjeva}.
According to Figure \ref{venus_range}, also in this case we have $|A_{\rm Rin} |\lesssim 10^{-13}$ m s$^{-2}$.
\begin{figure*}[ht!]
\centering
\begin{tabular}{cc}
\epsfig{file=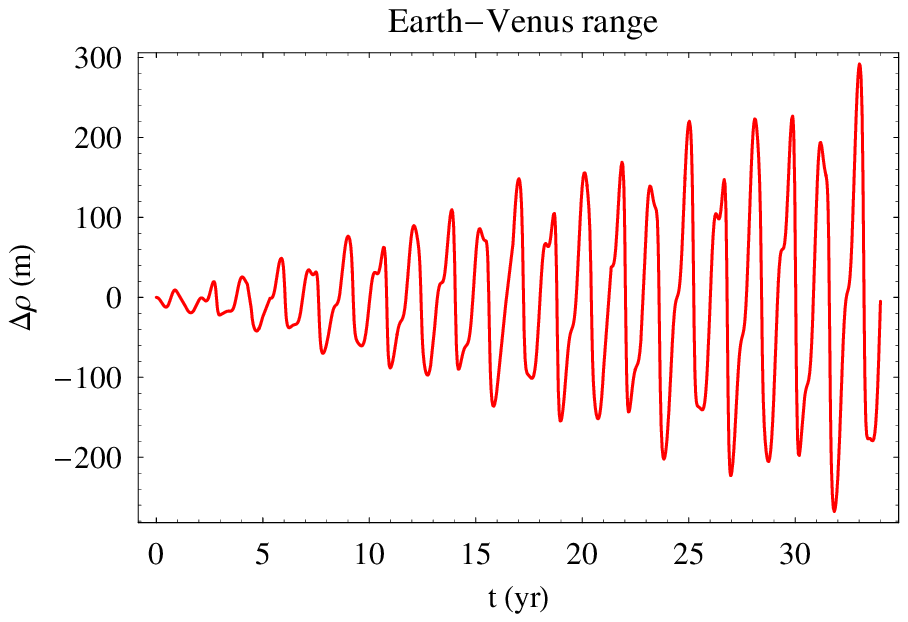,width=0.45\linewidth,clip=} &
\epsfig{file=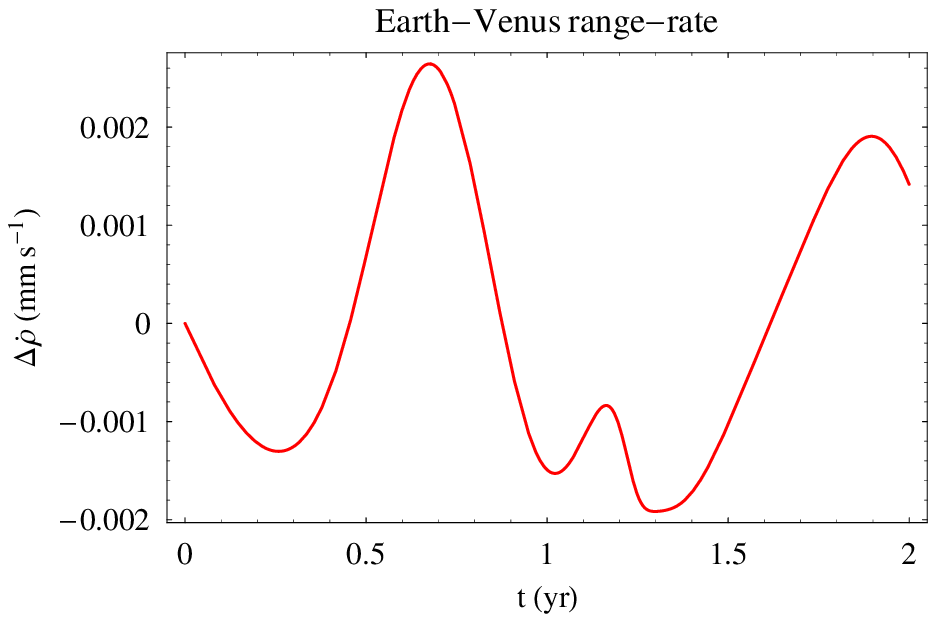,width=0.45\linewidth,clip=}
\end{tabular}
\caption{Left panel: Earth-Venus range perturbation $\Delta\rho$, in m, caused by $A_{\rm Rin} =2\times 10^{-13}$ m s$^{-2}$ over $\Delta t=34$ yr. Right panel: Earth-Venus (Magellan) range-rate perturbation $\Delta\dot\rho$, in mm s$^{-1}$, caused by $A_{\rm Rin} =2\times 10^{-13}$ m s$^{-2}$ over $\Delta t=2$ yr. The initial conditions have been retrieved from the NASA-JPL WEB interface HORIZONS.}\lb{venus_range}
\end{figure*}
{In Figure \ref{venus_rangeRATE_GTR} we depict the nominal Earth-Venus range-rate perturbation due GTR: cfr. with the right panel of Figure \ref{venus_range}. As anticipated in Section \ref{Rrange}, the different patterns are evident. }
\begin{figure*}[ht!]
\centering
\begin{tabular}{c}
\epsfig{file=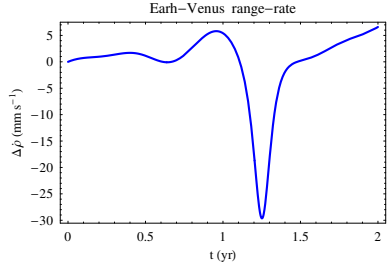,width=0.45\linewidth,clip=}
\end{tabular}
\caption{{Earth-Venus range-rate nominal perturbation $\Delta\dot\rho$, in mm s$^{-1}$, caused by GTR over $\Delta t=2$ yr. The initial conditions have been retrieved from the NASA-JPL WEB interface HORIZONS.}}\lb{venus_rangeRATE_GTR}
\end{figure*}

Tighter constraints come from Mars. According to Table 1 of  Ref.~\cite{pitjeva}, the 6-yr long range residuals of the Odyssey spacecraft (2002-2008) have a rms error of $1.2$ m; see also Figure B-11 of Ref.~\cite{folkner}. Figure \ref{mars_range} shows that, in this case, $A_{\rm Rin} $ is constrained at a $10^{-15}$ m s$^{-2}$ level.
\begin{figure*}[ht!]
\centering
\begin{tabular}{cc}
\epsfig{file=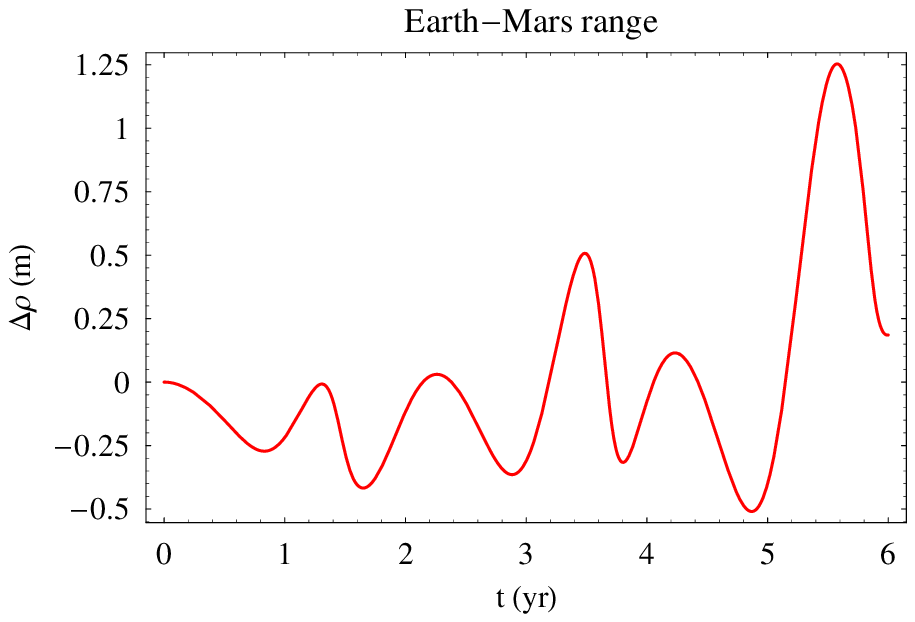,width=0.45\linewidth,clip=}&
\epsfig{file=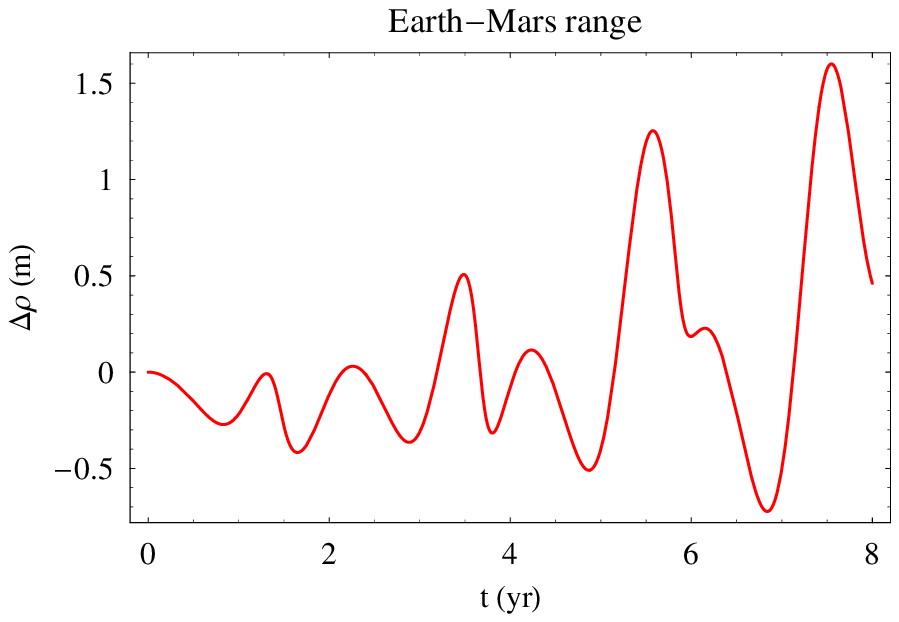,width=0.45\linewidth,clip=}
\end{tabular}
\caption{Left panel: Earth-Mars (Odyssey) range perturbation $\Delta\rho$, in m, caused by $A_{\rm Rin} =1\times 10^{-15}$ m s$^{-2}$ over $\Delta t=6$ yr. Right panel: Earth-Mars (MGS) range perturbation $\Delta\rho$, in m, caused by $A_{\rm Rin} =1\times 10^{-15}$ m s$^{-2}$ over $\Delta t=8$ yr. The initial conditions have been retrieved from the NASA-JPL WEB interface HORIZONS. }\lb{mars_range}
\end{figure*}
Similar results come from the range-residuals of the Mars Global Surveyor (MGS) spacecraft, covering 8 yr (1998-2006) and accurate to $1.4$ m \cite{pitjeva}; Figure \ref{mars_range} shows that they practically constrain $A_{\rm Rin} $ at the same level. It turns out that the range-rate residuals of Viking and Pathfinder \cite{pitjeva} do not yield constraints competitive with those from the range. {The pattern of the GTR signal is different with respect to the RIN one, as clearly shown by Fig. 23 of Ref. \cite{Iorio011} covering 5 yr.}

The present-day available range-residuals of the Cassini spacecraft orbiting Saturn cover about 2 yr (2004-2006) and are accurate to 20 m; see Table 1 of Ref.~\cite{pitjeva} and Figure B-20 of Ref.~\cite{DE}. They allow to constrain $A_{\rm Rin} $ at a $10^{-14}$ m s$^{-2}$ level, as shown by Figure \ref{saturn_range}.
\begin{figure*}[ht!]
\centering
\begin{tabular}{c}
\epsfig{file=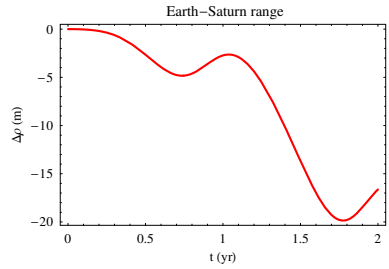,width=0.45\linewidth,clip=}
\end{tabular}
\caption{Earth-Saturn (Cassini) range perturbation $\Delta\rho$, in m, caused by $A_{\rm Rin} =1\times 10^{-14}$ m s$^{-2}$ over $\Delta t=2$ yr.  The initial conditions have been retrieved from the NASA-JPL WEB interface HORIZONS. }\lb{saturn_range}
\end{figure*}
{Also in this case, the Earth-Saturn range signal due to GTR is quite different with respect to the RIN one, as it can be noticed from Fig. 45 of Ref. \cite{Iorio011} covering 5 yr.}

At this point it is important to stress that our results clearly show that $A_{\rm Rin}$ cannot be the cause of the Pioneer anomaly. Indeed, $A_{\rm Rin}$ is a constant for a specific  system in consideration \cite{gru} in the sense that for a given central body of mass $M$ acting as source of the gravitational field $A_{\rm Rin}$ is fixed: neither spatial nor temporal variations are admitted. Thus, it is not possible that $A_{\rm Rin}$, being constrained to $10^{-14}-10^{-15}$ m s$^{-2}$ level by the inner planets, suddenly jumps to $10^{-10}$ m s$^{-2}$ in the regions in which the Pioneer anomaly manifested itself in its presently known form. \textcolor{black}{See also the discussion at the end of Section \ref{perieli} in which different empirical quantities are used to constrain $A_{\rm Pio}$  after 20 au.}

Moving to the neighborhood of the Earth, Figure B-1 of Ref.~\cite{DE} shows that the  residuals of the Earth-Moon range, constructed from the data continuously collected with the Lunar Laser Ranging (LLR) technique \cite{llr}, are at a cm-level since about 1990. Figure \ref{moon_range} depicts the  lunar range signature over $\Delta t=20$ yr caused by a terrestrial RIN $A_{\rm Rin} =5\times 10^{-16}$ m s$^{-2}$; its magnitude is as large as  4 cm at most, in agreement with the lunar laser ranging residuals. Larger values for $A_{\rm Rin} $ would yield signals too large; for example, it turns out that $A_{\rm Rin} =1\times 10^{-15}$ m s$^{-2}$ yields a RIN range signal of almost 10 cm.
\begin{figure*}[ht!]
\centering
\begin{tabular}{c}
\epsfig{file=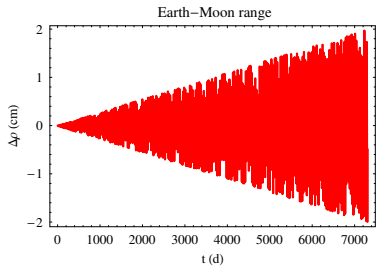,width=0.45\linewidth,clip=}
\end{tabular}
\caption{Earth-Moon  range perturbation $\Delta\rho$, in cm, caused by a terrestrial $A_{\rm Rin} =5\times 10^{-16}$ m s$^{-2}$ over $\Delta t=20$ yr.  The initial conditions have been retrieved from the NASA-JPL WEB interface HORIZONS. }\lb{moon_range}
\end{figure*}

It must be pointed out that, in principle, the bounds obtained so far might be somewhat too tight since the unmodeled RIN signatures, if they really existed in the data, may have been partially removed in the estimation of the initial conditions in the data reduction process; actually, RIN should be explicitly modeled, and a dedicated parameter accounting for it should be solved-for along with the other ones routinely estimated in the usual way.
First attempts towards the implementation of such a strategy applied to a generic extra-acceleration $A$ radially directed towards the Sun may have been recently performed by Folkner in Ref.~\cite{folkner}. He  obtained from the DE ephemerides by JPL an upper bound of $A\leq 1\times 10^{-14}$ m s$^{-2}$ on the Earth-Cassini range \cite{folkner}, in agreement with our Figure \ref{saturn_range}. On the other hand, the bound obtained from Mars in Ref.~\cite{folkner} is $A\leq 8\times 10^{-14}$ m s$^{-2}$, which is about one order of magnitude larger than ours in Figure \ref{mars_range}. However, it must be noted that, basically, no details have been released in Ref.~\cite{folkner} either concerning the  data set of the spacecrafts used for Mars nor about the methodology adopted to constrain $A$ from them.
\textcolor{black}{
\subsection{The perihelion precessions}\lb{perieli}
Another possible approach to constrain  $A_{\rm Rin}$ consists of looking at the empirically determined corrections $\Delta\dot\varpi$ to the standard Newtonian/Einsteinian secular precessions of the longitudes of the planetary perihelia $\varpi\doteq \Omega+\omega$, and compare them to theoretically predicted RIN-type precessions of $\varpi$ \cite{Sante}. \\ \\
The corrections $\Delta\dot\varpi$ have recently been estimated as solved-for parameters by independent teams of astronomers \cite{pitjeva,Fie010,FiengaJournees010} by fitting accurate dynamical force models of their ephemerides, which only include the usual Newtonian/Einsteinian dynamics, to long observational records spanning almost one century.
They are listed in Table \ref{perirate}.
\begin{table*}[ht!]
\caption{Estimated corrections $\Delta\dot\varpi$, in milliarcseconds per century (mas cty$^{-1}$), to the standard Newtonian/Einsteinian secular precessions of the longitudes of the perihelia $\varpi$ of the eight planets plus Pluto determined with the EPM2008 \protect{\cite{pitjeva}}, the INPOP08 \protect{\cite{Fie010}},  and the INPOP10a \protect{\cite{FiengaJournees010}} ephemerides. Only the usual Newtonian/Einsteinian dynamics was modelled, so that, in principle, the corrections $\Delta\dot\varpi$ account for any other unmodelled/mismodelled dynamical effect. Concerning the values quoted in the third column from the left, they correspond to the smallest uncertainties reported in Ref.~\protect{\cite{Fie010}}. Note the small uncertainty in the correction to the precession of the terrestrial perihelion, obtained by processing Jupiter VLBI data \protect{\cite{Fie010}}.
}\label{perirate}
\centering
\bigskip
\begin{tabular}{llll}
\hline\noalign{\smallskip}
Planet & $\Delta\dot\varpi$  \protect{\cite{pitjeva}}  & $\Delta\dot\varpi$  \protect{\cite{Fie010}} &  $\Delta\dot\varpi$  \protect{\cite{FiengaJournees010}} \\
\noalign{\smallskip}\hline\noalign{\smallskip}
Mercury & $ -4 \pm 5 $  & $ -10\pm 30$ & $ 0.2\pm 3$ \\
Venus & $ 24\pm 33$  & $-4\pm 6 $ & $ - $ \\
Earth & $ 6\pm 7$  & $ 0 \pm 0.016 $ & $ - $\\
Mars & $ -7\pm 7$  & $0\pm 0.2 $ & $ - $\\
Jupiter & $ 67\pm 93$  & $142\pm 156$ & $ - $\\
Saturn & $ -10\pm 15$ & $-10\pm 8$ & $ 0\pm 2$ \\
Uranus & $ -3890\pm 3900$  & $0\pm 20000$ & $ - $\\
Neptune & $ -4440\pm 5400 $  & $0\pm 20000$ & $ - $\\
Pluto & $ 2840 \pm 4510 $  & $-$ & $ - $\\
\noalign{\smallskip}\hline\noalign{\smallskip}
\end{tabular}
\end{table*}
\\ \\
The secular precession of $\varpi$ caused by a constant and uniform  extra-acceleration $A$ radially directed towards the Sun has been computed by the authors of Refs.~\cite{Pio1,Pio2,Pio3} in the framework of the investigations of the Pioneer anomaly. It is
\eqi\dert{\varpi}{t}=-A\rp{\sqrt{1-e^2}}{na}\lb{peripalla}.\eqf \\ \\
By applying \rfr{peripalla} to RIN, a comparison of it with the most accurate values of $\Delta\dot\varpi$ in Table \ref{perirate} yields the upper bounds on $A_{\rm Rin}$ listed in Table \ref{cazzarola}.
\begin{table*}[ht!]
\caption{Upper bounds on the magnitude of RIN, in m s$^{-2}$, from a comparison of the theoretical prediction of \rfr{peripalla} for the RIN-induced secular precession of the longitude of the pericenter $\varpi$ of a test particle and the most accurate values of the empirically determined corrections $\Delta\dot\varpi$ to the standard Newtonian/Einsteinian secular planetary precessions of the perihelia quoted in Table \ref{perirate}.
}\label{cazzarola}
\centering
\bigskip
\begin{tabular}{ll}
\hline\noalign{\smallskip}
Planet &  $ A_{\rm Rin}$ (m s$^{-2}$)\\
\noalign{\smallskip}\hline\noalign{\smallskip}
Mercury & $ \leq 2\times 10^{-13} $  \\
Venus & $ \leq 3\times 10^{-13}$   \\
Earth & $\leq 7\times 10^{-16} $  \\
Mars & $\leq 7\times 10^{-15}$  \\
Jupiter & $\leq 2\times 10^{-12}$  \\
Saturn & $\leq 3\times 10^{-14}$   \\
Uranus & $\leq 4\times 10^{-11}$  \\
Neptune & $\leq 4\times 10^{-11} $  \\
Pluto & $\leq 3\times 10^{-11} $\\
\noalign{\smallskip}\hline\noalign{\smallskip}
\end{tabular}
\end{table*}
Remarkably, the constraints on $A_{\rm Rin}$ of Table \ref{cazzarola}, obtained from the planetary perihelia, are in substantial agreement with those of Section \ref{rangi} inferred from the range and range$-$rate residuals. The perihelion of the Earth yields a bound as low as $A_{\rm Rin}\leq 7\times 10^{-16}$ m s$^{-2}$.\\ \\
Incidentally, it can also be noticed from Table \ref{cazzarola} that the largest admissible values for $A_{\rm Rin}$ from the perihelia of Uranus, Neptune and Pluto are smaller than the smallest value of the anomalous Pioneer acceleration, i.e. $A_{\rm Pio} = 7.41\times 10^{-10}$ m s$^{-2}$ ($1\sigma-$level), by a factor $18-25$, respectively. At $3\sigma-$level, i.e. for $A_{\rm Pio} = 4.75\times 10^{-10}$ m s$^{-2}$, they are smaller than it by a factor $12-16$.}
\section{Summary and conclusions}\lb{conclusioni}
We analytically worked out the perturbing effects which a Rindler-type anomalous acceleration $A_{\rm Rin}$, naturally arising from a 2-dimensional dilaton-based long-range modification of gravity, would induce on the orbital motion of a test particle orbiting a central body acting as source of the modified gravitational field. In particular, in view of a comparison with the most recent observations we focussed on the effects of $A_{\rm Rin}$ on the two-body range $\rho$ and range-rate $\dot\rho$ because they are direct, unambiguous observables widely used in satellite and planetary investigations.

It turns out that  $A_{\rm Rin}$ does actually affect $\rho$ and $\dot \rho$ with long-term signatures which can fruitfully be compared with the residuals for such observables built by processing extended data records for some planets and the Moon with the latest ephemerides in which the Rindler perturbation has not been explicitly modeled. It turns out that Mercury and Venus constrain a solar $A_{\rm Rin}$ to a level of about $10^{-13}$ m s$^{-2}$. The bounds inferred from recent data sets of some spacecrafts orbiting Mars are of the order of $10^{-15}$ m s$^{-2}$, while the Cassini range residuals yield $A_{\rm Rin}\lesssim 1\times 10^{-14}$ m s$^{-2}$ at the Saturn's distance. \textcolor{black}{Analogous constraints are obtained by using the empirically determined corrections $\Delta\dot\varpi$ to the standard Newtonian/Einsteinian secular precessions of the planetary perihelia: the Earth yields a tighter bound as low as $A_{\rm Rin}\lesssim 7\times 10^{-16}$ m s$^{-2}$}. The tightest constraints come from the lunar range residuals according to which $A_{\rm Rin}\lesssim 5\times 10^{-16}$ m s$^{-2}$ for a modification of the Earth's gravitational field. Another analysis existing in literature, likely based on a different method, points towards $A_{\rm Rin}\leq 8\times 10^{-14}$ m s$^{-2}$ at the Mars' distance, while it agrees with our result for Saturn.

In principle, $A_{\rm Rin}$ should be explicitly modeled, the entire planetary and satellite data sets used should be re-processed with such modified dynamical models, and a dedicated solve-for parameter should be estimated. Such an approach is certainly rather expensive in terms of computational burden and time required, especially if other, more conventional tasks are pressing and have to be mandatorily performed. Moreover, if one is interested in other exotic  effects predicted by some different modified gravities the entire procedure has to be repeated with the new model.
Instead, the approach followed here in the case of the Rindler-type modification of gravity at large distances can be  easily and straightforwardly extended to other long-range modified models of gravity as well


\end{document}